\documentstyle[aps,prl,multicol,epsf]{revtex}
\title{\bf Extremal properties of random trees}
\author{E.~Ben-Naim$^1$, P.~L.~Krapivsky$^{2}$, Satya N. Majumdar$^{3,4}$}
\address{$^1$Theoretical Division and Center for Nonlinear Studies, 
Los Alamos National Laboratory, Los Alamos, NM 87545, USA}
\address{$^2$Center for Polymer Studies and
Department of Physics, Boston University, Boston, MA 02215, USA}
\address{$^3$CNRS, IRSAMC, Laboratoire de Physique Quantique,
Universite' Paul Sabatier, 31062 Toulouse, France}
\address{$^4$Tata Institute of Fundamental Research, Homi Bhabba Road,
Mumbai-400005, India}

\begin{document}
\maketitle
\begin{abstract}

\noindent
We investigate extremal statistical properties such as the maximal and
the minimal heights of randomly generated binary trees.  By analyzing
the master evolution equations we show that the cumulative
distribution of extremal heights approaches a traveling wave form.
The wave front in the minimal case is governed by the
small-extremal-height tail of the distribution, and conversely, the
front in the maximal case is governed by the large-extremal-height
tail of the distribution.  We determine several statistical
characteristics of the extremal height distribution analytically.  In 
particular, the expected minimal and maximal heights grow
logarithmically with the tree size, $N$, $h_{\rm min} \sim v_{\rm
  min}\ln N$, and $h_{\rm max}\sim v_{\rm max}\ln N$, with $v_{\rm
  min}=0.373365\ldots$ and $v_{\rm max}=4.31107\ldots$, respectively.
Corrections to this asymptotic behavior are of order ${\cal O}(\ln \ln
N)$.

\end{abstract}

\begin{multicols}{2}

Random trees play an important role in data storage and retrieval
algorithms in computer science \cite{robson,pittel,Dev,pn,mahm,knuth}.
They also arise in physical situations such as collision 
processes in gases\cite{van}, random fragmentation processes
\cite{HO,PS}, and diffusion-limited aggregation\cite{DLA}. 
In each case, extremal characteristics such as the
maximal or the minimal height of the tree, namely, the maximal
\cite{Dev} or the minimal \cite{pn} number of bonds separating the
tree root from a node are of interest.  In data storage algorithms,
these distances yield the best-case or the worst-case-scenario
performances.  In kinetic theory, the largest Lyapunov exponent is
related to the maximum height problem\cite{van}.

In this article, we study extremal properties of randomly generated
binary trees using rate equation theory. Techniques developed in
aggregation processes are well suited for treating random trees since
the tree merger process is simply an aggregation process.  We study
the distributions of extremal (both minimal and maximal) heights of a
tree.  In both cases, the average extremal tree height grows
logarithmically with the number of leaves and the cumulative
distribution of extremal tree heights approaches a traveling wave
solution. The logarithmic growth prefactors equal the traveling wave
velocities, which are set by a velocity selection principle.
These velocities can be alternatively obtained using a simpler (and
independent) intuitive argument.  Interestingly, the wave front in the
minimal case is determined by the small-height tail of the
distribution, while in the maximal case it is determined by the
large-height tail of the distribution.

Let us introduce the tree generation model. Initially, the system
consists of an infinite number of trivial (single-leaf) trees.  Then,
two trees are picked at random and attached to a common root.  This
merging process is repeated indefinitely with rate set to 2 without
loss of generality.  Let $c(t)$ be the number density of trees at time
$t$.  Initially, $c(0)=1$, and since this quantity evolves according
to $dc/dt=-c^2$ one has
\begin{equation}
\label{ct}
c(t)={1\over 1+t}. 
\end{equation}
Mass conservation implies that $N$, the average number of leaves in a
tree, grows linearly with time, $N=c^{-1}=1+t$. While the corresponding
mass distribution has been extensively studied in coalescence
processes\cite{mvs,Aldous}, we are interested here in the distribution
of extremal characteristics such as the minimal and maximal number of
bonds between the tree root and its nodes.  The leading behavior of 
the average of these distributions can be obtained from the following 
intuitive argument.

The distribution of tree heights, namely, of the distances between the
tree root and the nodes can be obtained immediately.  Let 
$P_n(t)$ be the probability that the distance between a randomly chosen
leaf and the root of the parent tree equals $n$ at time $t$. As the
tree generation process is random, $P_n(t)$ obeys Poisson statistics
\begin{equation}
\label{pn}
P_n(t)={[h(t)]^n\over n!}e^{-h(t)},
\end{equation}
with $h(t)$ the average tree height.  Consider a leaf in the system.  Each
time its corresponding tree merges with another tree, the distance to the
root is augmented by one. This process occurs with rate 2 and hence,
$dh/dt=2c$. Integrating this equation subject to the initial condition
$h(0)=0$ yields $h(t)=2\ln (1+t)=2\ln N$.  One anticipates that the expected
minimal number grows logarithmically as well, $h_{\rm min}\simeq v_{\rm
  min}\ln N$.  To estimate $h_{\rm min}$ we sum the small-$n$ tail of the
normalized height distribution $c^{-1}P_n$ to unity, $\sum_{n=0}^{h_{\rm
    min}}c^{-1}P_n=1$.  Substituting Eqs.~(\ref{ct})--(\ref{pn}), the relation
$h_{\rm min}\simeq v_{\rm min}\ln N$, and the Stirling formula $\ln n!~\sim
n\ln n-n$ into the above relation we obtain the transcendental equation
\begin{equation}
\label{veq}
v\ln {2e\over v}=1.
\end{equation}
This equation has two solutions with the lower (higher) velocity
corresponding to the growth of the average minimal (maximal) tree height.
Indeed, repeating the above steps for the maximal height using $\sum_{
n=h_{\rm max}}^{\infty} c^{-1}P_n=1$ again leads to the same equation.
Solving Eq.~(\ref{veq}) yields 
\begin{eqnarray}
\label{nv}
h_{\rm min}&\simeq v_{\rm min}\ln N,
\quad\qquad v_{\rm min}&=0.373365;\\
h_{\rm max}&\simeq v_{\rm max}\ln N,
\quad\qquad v_{\rm max}&=4.31107.\nonumber
\end{eqnarray}
This probabilistic argument correctly predicts both velocities, and
additionally, it demonstrates that the two extremal statistics are
intimately related.  We note that the latter maximal height value has
emerged from quite different calculations in studies of collision
processes in gases \cite{van} and fragmentation processes\cite{HO,PS}.

We now turn to studying the entire distribution of extremal
characteristics. Rather than considering the two extremal height
distributions separately, we study a more general model which
interpolates between the two cases.  In this model, each tree carries
an extremal height $k$.  The result of a merger between trees with
extremal heights $k_1$ and $k_2$, is a new tree with extremal height
$k$ given by
\begin{equation}
\label{min}
k=\cases{ {\rm min}(k_1, k_2)+1 & with prob.~ $p$, \cr
{\rm max}(k_1,k_2)+1 & with prob.~ $1-p$ .\cr}
\end{equation}
Here, $p$ is a mixing parameter whose limits $p=1$ and $p=0$ correspond
to the minimal and the maximal heights problems, respectively.   

The number density of trees with extremal height $k$, $c_k(t)$,
evolves according to the master equation
\begin{equation}
\label{ckmin}
{d c_k\over dt}=c_{k-1}^2\!\!-2cc_k\!+\!2pc_{k-1}
\!\!\sum_{j=k}^\infty c_j\!+\!2(1\!-\!p)c_{k-1}\!\!\sum_{j=0}^{k-2}c_j.
\end{equation}
Here $c=\sum_{j\geq 0}c_j$ is the total tree density and one can
verify that it indeed evolves according to $dc/dt=-c^2$.  The master
equation (\ref{ckmin}) should be solved subject to the initial
condition $c_k(0)=\delta_{k,0}$.  It proves useful to introduce the
cumulative fractions
\begin{equation}
\label{Ak}
A_k=c^{-1}\sum_{j=k}^\infty c_j,
\end{equation}
and a new time variable 
\begin{equation}
\label{T}
T=\int_0^t d\tau\,c(\tau)=\ln(1+t).
\end{equation}
These variables recast Eqs.~(\ref{ckmin}) into
\begin{equation}
\label{Akmin}
{dA_k\over dT}=-A_k+2(1-p)A_{k-1}+(2p-1)A_{k-1}^2,
\end{equation}
which should be solved subject to the step function initial
conditions, $A_k(0)=1$ for $k\leq 0$ and $A_k(0)=0$ otherwise.

In the long time limit, $A_k(T)$ approaches a traveling wave form,
$A_k(T)\to A(k-vT)$, with $A(x)$ being a solution of the nonlinear
difference-differential equation
\begin{equation}
\label{Amin}
vA'(x)=A(x)\!-\!2(1\!-\!p)A(x\!-\!1)\!-\!(2p\!-\!1)A^2(x\!-\!1),
\end{equation}
subject to the boundary conditions $A(-\infty)=1$ and $A(\infty)=0$.
Fortunately, the velocity $v$ can be determined without solving the
nonlinear nonlocal equation (\ref{Amin}) exactly. To determine $v$, it
is enough to analyze the asymptotic behavior of the front at one of
its two tails.  Different considerations apply in the regions $p\leq
1/2$ and $p\geq 1/2$. We first consider the case $1/2\leq p \leq
1$. Here, Eq.~(\ref{Amin}) admits an exponential solution in
the small-$k$ tail, $A(x)\to 1-e^{\lambda x}$ as $x\to -\infty$.
Substituting this form in Eq.~(\ref{Amin}), we find that the yet to be
determined velocity $v$ and decay exponent $\lambda$ are related via
\begin{equation}
\label{v}
v={1-2pe^{-\lambda}\over \lambda}.
\end{equation}
While a class of velocities is in principle possible, the extremum
value is selected for compact initial conditions.  This behavior is
similar to velocity selection occurring for example in the classic Fisher
reaction-diffusion equation \cite{Murray,Bram}.
Evaluating this extremum yields a generalization of the transcendental
equation (\ref{veq}) 
\begin{equation}
\label{vp1}
v\ln {2ep\over v}=1, \qquad \qquad p>{1\over 2}.
\end{equation}
In particular, $v\to 1$ when $p\to 1/2$. In the minimal height case
($p=1$) we recover the aforementioned value $v_{\rm min}= 0.373365$.
The selected decay coefficient satisfies $2p(1+\lambda)=e^{\lambda}$
and in the minimal case $\lambda= 1.67835$.

Let us now turn to the complementary $p<1/2$ case where contrary to
the $p>1/2$ case, the large-extremal-height tail admits an
exponentially decaying solution $A(x)=e^{-\mu x}$, as $x\to +\infty$.
Here, the yet to be determined velocity and decay coefficient are
related via
\begin{equation}
\label{vf}
v={ {2(1-p)e^{\mu}-1}\over {\mu}}.
\end{equation}
Again, applying the velocity selection principle implies that the 
minimal possible velocity is selected. The selected decay coefficient 
satisfies $2(1-p)(1-\mu)=e^{-\mu}$ and the selected velocity obeys 
\begin{equation}
\label{vp2}
v\ln {2e(1-p)\over v}=1, \qquad \qquad p<{1\over 2}.
\end{equation}
In the maximal case ($p=0$) one recovers the velocity $v_{\rm
  max}= 4.31107$ and additionally, the decay coefficient is
$\mu=0.768039$. While we have not proved this selection principle, our
numerical integration of Eq.~(\ref{Akmin}) supports the findings in
the minimal and the maximal cases. We confirmed that a traveling wave
solution is indeed approached, and that the selected velocities fall
within 0.1\% of the theoretical values.  Curiously, this is a unique
case where velocity selection is also supported by an independent
physical  argument.

Interestingly, different mechanisms drive the front in the
regions $p<1/2$ and $p>1/2$. In the region $p>1/2$ which includes the
minimal case, the small-extremal-height tail of the distribution
dictates the velocity and in fact, the entire distribution is enslaved
to this exponential tail. The opposite is true for $p<1/2$ which
includes the maximal case. Here, the large-extremal-height tail of the
distribution governs the wave velocity and the wave form.  These
behaviors are physical, especially when the limiting cases are
considered: the distribution of minimal (maximal) tree heights is
governed by extremely small (large) fluctuations.  The point $p_c=1/2$
can be regarded as a critical point.  In particular, the different
velocity equations (\ref{vp1}) and (\ref{vp2}) imply non-analytic
behavior at $p_c=1/2$ where $v_c=1$, and analysis of the leading
behavior in the vicinity of this point yields
\begin{equation}
\label{pc}
|v-v_c|\simeq 2|p-p_c|^{1/2},\quad{\rm when}\quad p\to p_c.
\end{equation}
Exact analysis of the special case $p=1/2$ is given below.  

Asymptotically, while the wave front advances at a constant rate $v$, there
is a slow $T^{-1}$ correction in the leading order, resulting in a
logarithmic correction to the front position.  A similar correction was first
derived by Bramson in the context of reaction-diffusion equations
\cite{Bram}, and was subsequently generalized \cite{vS,B+D,EvS}.  We now
calculate the leading correction employing the approach of Ref.\cite{B+D}.
Let us first consider the $p>1/2$ case.  Substituting $A_k(T)=1-a_k(T)$ into
Eq.~(\ref{Akmin}), and ignoring the quadratic term $a_{k-1}^2$, yields
\begin{equation}
\label{Bk}
{da_k\over dT}=-a_k+2pa_{k-1}.
\end{equation}
Substituting the scaling solution 
\begin{equation}
\label{Bkscal}
a_k(T)=T^\alpha G(x\,T^{-\alpha})e^{\lambda x},\quad x=k-vT-w(T).
\end{equation}
into (\ref{Bk}) shows that different leading orders are compatible
provided that the exponent $\alpha=1/2$ and that the correction to the
front location is $w(T)=\beta \ln T$.  The former constraint reflects
a hidden diffusive scale and the latter gives the aforementioned
logarithmic correction with yet undetermined amplitude
$\beta$. Substituting these behaviors in Eq.~(\ref{Bk}) we get
\begin{eqnarray*}
0&=&T^{1/2}\left(2pe^{-\lambda}-1+\lambda v\right)G(z)
+\left(v-2pe^{-\lambda}\right)G'(z)\\
&+&T^{-1/2}\left[e^{-\lambda}G''(z)+{1\over 2}\,zG'(z)
+{2\lambda \beta-1\over 2}\,G(z)\right]+\ldots
\end{eqnarray*}
The terms of the leading ${\cal O}(T^{1/2})$ order cancel when
$2pe^{-\lambda}+\lambda v=1$, i.e., when the velocity $v$ and the
decay exponent $\lambda$ are related through Eq.~(\ref{v}). The terms
of order ${\cal O}(1)$ cancel when $2pe^{-\lambda}=v$.  Remarkably,
this relation together with Eq.~(\ref{v}) hold only in the extremal
point where $v$ and $\lambda$ are given by Eqs.~(\ref{vp1}) and
(\ref{v}), respectively. Finally, the terms in the lowest ${\cal
O}(T^{-1/2})$ order cancel when the scaling function $G(z)$ satisfies
the parabolic cylinder equation
\begin{equation}
v{d^2G\over dz^2}+ z\,{dG\over dz} +(2\beta \lambda-1)G(z)=0.
\label{para}
\end{equation}
This differential equation should be solved subject to the appropriate
boundary conditions: (i) $G(z)\to 0$ for $z\to -\infty$ as $B_k(T)$
must vanish when $T\to\infty$, and (ii) $G(z)\sim z$ for $z\to 0$ to
ensure that $B_k(T)$ is independent of $k$ for large $k$. The former
boundary condition selects one of the two possible solutions,
$G(z)=C\,e^{-y^2/4}\,D_\nu(y)$, where $y=z/\sqrt{v}$,
$\nu=2(\beta \lambda-1)$, and $D_\nu(y)$ is the parabolic cylinder
function with index $\nu$.  The second boundary condition fixes the
index, $\nu=1$, implying $\beta=3/(2\lambda)$.  Hence, we find that
the expected extremal tree height, $\langle k\rangle=c^{-1}\sum_k kc_k$,
grows with $N$ as
\begin{equation}
\label{xmin}
\langle k\rangle= v\ln N+{3\over 2\lambda}\,\ln \ln N,
\end{equation}
where the relation $T=\ln N$ was used. Obviously, the latter log-log
correction can not be obtained from the heuristic argument presented
earlier. The same analysis can be carried out for the $p<1/2$ case
where we find,
\begin{equation}
\label{xmax}
\langle k\rangle=v\ln N-{3\over 2\mu}\,\ln \ln N,
\end{equation}
with $v$ and $\mu$ given by Eqs.~(\ref{vp2}), and (\ref{vf}),
respectively.  For completeness, we merely quote results of a more
sophisticated approach \cite{EvS} which allows calculation of the
second leading correction to the growth rate
\begin{equation}
{d\langle k\rangle\over dT}=\cases{{3\over 2\lambda}\,T^{-1}
+{3\over {\lambda}^2}\,\sqrt{\pi\over 2v}\,T^{-3/2},
& $p>1/2$, \cr
-{3\over 2\mu}\,T^{-1}
+{3\over {\mu}^2}\,\sqrt{\pi\over 2v}\,T^{-3/2},
& $p<1/2$.\cr}
\end{equation}

We now return to the critical case $p=1/2$. The critical behavior is
simpler because Eq.~(\ref{Akmin}) becomes linear, and it can be easily
solved by the generating function method.  {}From Eq.~(\ref{Akmin}) we
find that the generating function $Q(z,T)=\sum_{k\geq 1}A_k z^k$
satisfies the differential equation $dQ/dT=z+(z-1)Q(z,T)$, subject to
the initial condition $Q(z,0)=0$. This equation admits the solution
$Q(z,T)=z[1-e^{-(1-z)T}]/(1-z)$, and expanding in powers of $z$ gives
the cumulative distribution
\begin{equation}
A_k(T)=e^{-T}\sum_{m=k}^{\infty}{T^m\over {m!}}.
\label{aktc}
\end{equation}
Using Eq.~(\ref{Ak}) and $c(T)=e^{-T}$, we find that the extremal height
distribution is proportional to a Poisson distribution, $c_k(T)=e^{-2T}T^k/
k!$. Asymptotically, the normalized height distribution approaches a
Gaussian, \hbox{$c^{-1}c_k(T)\to \exp[-(k-T)^2/2T]/\sqrt{2\pi T }$}, and the
cumulative fractions easily follow
\begin{equation}
A_k(T)\to {1\over 2}{\rm  erf}\left[{k-T\over \sqrt{2T}}\right].
\end{equation}
Hence, both tails are Gaussian,
consistent with the fact that $\lambda=\mu=0$ when $p=1/2$.
Interestingly, the hidden diffusive scale becomes pronounced,
and the wave front broadens indefinitely with a width of the order 
$\sqrt{T}$.

Thus far, we have considered closed systems where only trivial trees are
initially present. However, in many applications such as data storage
algorithms, as well as in physical situations such as river
networks\cite{D+R} and fragmentation processes\cite{bk}, there may be a
constant input into the system. We therefore consider the natural case where
an initially empty system is subject to a uniform input of trivial trees. In
this case, we must add an additional input term $\delta_{k,0}$ into the
right-hand side of Eq.~(\ref{ckmin}). The initial condition now reads
$c_k(0)=0$. The overall density evolves according to $dc/dt=1-c^2$, i.e.,
$c(t)=\tanh(t)$. Hence, the system eventually reaches a steady state with
density $c=1$.

We restrict our attention to the steady state distributions which can be
obtained by equating the time derivatives in the master equations
(\ref{ckmin}) to zero. Again, we introduce the cumulative densities
$B_k=\sum_{j= k}^{\infty} c_j$ which at the steady state satisfy $B_0=c=1$
and
\begin{equation}
\label{bk}
B_k=(1-p)B_{k-1} +(p-1/2)B_{k-1}^2
\end{equation}
for $k\geq 1$.  Three different behaviors arise depending on whether
$p=1$, $0<p<1$, or $p=0$. For $p=1$, solving (\ref{bk}) recursively
gives $B_k=2^{-(2^k-1)}$. Therefore,
$c_k=2^{-(2^k-1)}\left(1-2^{-2^k}\right)$, implying that the minimal
height distribution decays as an unusual double-exponential.  For
$0<p<1$, one can neglect the nonlinear term in Eq.~(\ref{bk}) in the
large $k$ limit. Thence, $B_k\sim (1-p)^{k}$ implying a generic
exponential decay of the distribution $c_k$ at large $k$. The critical
behavior disappears and the only notable feature of the $p=1/2$ case
is that it is exactly solvable, $B_k=2^{-k}$.  For the maximal height
problem ($p=0$), the recursion (\ref{bk}) simplifies (in the large $k$
limit) to a differential equation $dB/dk=-B^2/2$ which is solved to
give $B_k\simeq 2k^{-1}$. Thus, the maximal height distribution
exhibits a power-law decay, $c_k\simeq 2k^{-2}$ for large $k$. To
summarize, we quote (up to a numeric prefactor) the three leading
large-$k$ behaviors
\begin{equation}
\label{ckinfmax}
c_k\sim\cases{
2^{-2^k},&$p=1$;\cr
(1-p)^k,&$0<p<1$;\cr
k^{-2},&$p=0$.\cr}
\end{equation}
Thus, input dramatically alters the height distributions
with a wide array of possible outcomes including double exponential,
exponential, and algebraic decays. 

In summary, we have shown that extremal properties of random trees can
be obtained by analyzing the corresponding nonlinear evolution
equations. The cumulative distributions of extremal tree heights
approach a traveling wave solution.  The mean extremal values grow
logarithmically with the tree size and there is an additional weak
double logarithmic correction. The corresponding growth velocities were
obtained from an elementary probabilistic argument and from
an extremum selection criteria on the traveling wave solution.
Interestingly, while the traveling wave velocity and form in the
minimal case is determined by extremely small height fluctuations, the
opposite holds for the maximal case. The transition between these two
behaviors is marked by a sharp phase transition in a model which
interpolates between the two extremal characteristics. Additionally,
we have showed that the presence of input may lead to double
exponential, exponential, or even algebraic decays of the extremal
height distribution. It will be interesting to apply rate equation
theory to a closely related random tree characteristics, e.g., the
bifurcation ratio and the rank\cite{DLA,D+R}.

\smallskip We are thankful to M.~B.~Hastings and Z.~Toroczkai for
interesting discussions, and to the 
DOE (W-7405-ENG-36) and NSF(DMR9978902) for
financial support.

\end{multicols}

\begin{references}


\bibitem{robson}  J.~M.~Robson,
                  Austr. Comput. J.  {\bf 11}, 151 (1979).

\bibitem{pittel}  B.~Pittel,
                  J. Math. Anal. Appl. {\bf 103}, 461 (1984).

\bibitem{Dev}     L.~Devroye,
                  J. ACM {\bf 33}, 489 (1986).

\bibitem{pn}      P.~Noguiera, Disc. Appl. Math. {\bf 109}, 253 (2001).

\bibitem{mahm}    H.~M.~Mahmoud, {\em Evolution of Random
                  Search Trees} (John Wiley, New York, 1992).

\bibitem{knuth}   D.~E.~Knuth, {\em The Art of Computer Programming,
                  vol. 3, Sorting and Searching}
                  (Addison-Wesley, Reading, 1998).

\bibitem{van}     R. van Zon, H. van Beijeren, and Ch. Dellago,
                  Phys. Rev. Lett. {\bf 80}, 2035 (1998).

\bibitem{HO}      T. Hattori and H. Ochiai, preprint (1998).

\bibitem{PS}      P.~L.~Krapivsky and S.~N.~Majumdar,
                  Phys. Rev. Lett. {\bf 85}, 5492 (2000).

\bibitem{DLA}     I.~Yekutieli and B.~B.~Mandelbrot, 
                  J.\ Phys.\ A {\bf 27}, 285 (1994);
                  T.~C.~Halsey, Europhys.\ Lett. {\bf 39}, 43 (1997).

\bibitem{mvs}     M.~V.~Smoluchowski, Z. Phys.\ Chem. {\bf 92}, 215 (1917).

\bibitem{Aldous}  For a recent review, see 
                  D.~Aldous, Bernoulli {\bf 5}, 3 (1999).

\bibitem{Murray}  J.~D.~Murray, {\em Mathematical Biology}
                  (Springer-Verlag, New York, 1989).

\bibitem{Bram}    M.~Bramson, {\em Convergence of Solutions of the
                  Kolmogorov Equation to Travelling Waves} (American
                  Mathematical Society, Providence, R.I., 1983).

\bibitem{vS}      W.~van~Saarloos,
                  Phys. Rev. A {\bf 39}, 6367 (1989).

\bibitem{B+D}     E.~Brunet and B.~Derrida,
                  Phys. Rev. E {\bf 56}, 2597 (1997).

\bibitem{EvS}     U.~Ebert and W.~van~Saarloos,
                  Phys. Rev. Lett. {\bf 80}, 1650 (1998);
                  Physica D {\bf 146}, 1 (2000).

\bibitem{D+R}     P.~S.~Dodds and D.~H.~Rothman,
                  Phys. Rev. E {\bf 59}, 4865 (1999).

\bibitem{bk}      E.~Ben-Naim and P.~L.~Krapivsky, 
                  Phys. Lett. A {\bf 275}, 48 (2000).

\end{references}
\end{document}